\documentclass[prd,preprint,superscriptaddress,showpacs,byrevtex]{revtex4}
\usepackage{bm}
\def\fsl#1{\setbox0=\hbox{$#1$}           
   \dimen0=\wd0                                 
   \setbox1=\hbox{/} \dimen1=\wd1               
   \ifdim\dimen0>\dimen1                        
      \rlap{\hbox to \dimen0{\hfil/\hfil}}      
      #1                                        
   \else                                        
      \rlap{\hbox to \dimen1{\hfil$#1$\hfil}}   
      /                                         
   \fi}                                         %
\newcommand{\be}{\begin{equation}}
\newcommand{\ee}{\end{equation}}
\newcommand{\bea}{\begin{eqnarray}}
\newcommand{\eea}{\end{eqnarray}}
\newcommand{\beq}{\begin{equation}}
\newcommand{\eeq}{\end{equation}}
\newcommand{\beqs}{\begin{eqnarray}}
\newcommand{\eeqs}{\end{eqnarray}}

\newcommand{\gsim}{\mathrel{\raisebox{-
.6ex}{$\stackrel{\textstyle>}{\sim}$}}}

\newcommand{\pslash}{p\hspace{-0.067in}\slash}

\newcommand{\kslash}{k\hspace{-0.067in}\slash}

\newcommand{\dslash}{D\hspace{-0.067in}\slash}
\begin{document}
\title{ Proof of Factorization of Heavy Quarkonium Production in Non-Equilibrium QCD at RHIC and LHC }
\author{Gouranga C Nayak }\thanks{G. C. Nayak was affiliated with C. N. Yang Institute for Theoretical Physics in 2004-2007.}
\affiliation{ C. N. Yang Institute for Theoretical Physics, Stony Brook University, Stony Brook NY, 11794-3840 USA}
%
\begin{abstract}
Recently we have proved the NRQCD factorization in heavy quarkonium production at high energy colliders at all orders in coupling
constant. In this paper we extend this to non-equilibrium QCD and prove the NRQCD factorization in heavy quarkonium
production at all orders in coupling constant in non-equilibrium QCD. This proof is necessary to study heavy
quarkonium production from quark gluon plasma at RHIC and LHC.
\end{abstract}
\maketitle
\pagestyle{plain}
\pagenumbering{arabic}
\section{Introduction}
Recently we have proved the non relativistic QCD (NRQCD) factorization in heavy quarkonium production at high energy colliders at all orders in coupling
constant in \cite{nayaknr}. In this paper we extend this to non-equilibrium QCD and prove the NRQCD factorization at
all orders in coupling constant in non-equilibrium QCD which is necessary to study quark-gluon plasma at RHIC and LHC.

Just after the big bang our universe was filled with a state of matter known as quark-gluon plasma.
The temperature of quark-gluon plasma is $\gsim$ 200 MeV which is much larger than the
temperature of the sun. Besides black holes, the quark-gluon plasma is denser than all
other forms of the matter we know so far. Hence recreating this
early universe scenario in the laboratory is challenging. At present RHIC (Au-Au collisions at
$\sqrt{s}_{NN}$ = 200 GeV) and LHC (Pb-Pb collisions at $\sqrt{s}_{NN}$ = 2.76 TeV in the run-1)
provide the best facilities to study the production of quark-gluon plasma in the laboratory.
In the run-2 the LHC collides two lead nuclei at $\sqrt{s}_{NN}$ = 5.02 TeV which
creates even more energy density, {\it i. e.}, it produces quark-gluon plasma with higher
temperature.

The main difficulty we face at high energy heavy-ion colliders is to detect the quark-gluon plasma and to study
its properties. This is because we have not seen quarks and gluons directly. Hence the detection of
quark-gluon plasma has to be done by using indirect signatures.
Heavy quarkonium is an useful probe to study quark-gluon plasma.
$J/\psi$ suppression is suggested to be one of the main signature of the quark-gluon plasma
detection \cite{satz}. This is based on the argument of Debye screening in
quark-gluon plasma. At high temperature the Debye screening length becomes smaller than the
$j/\psi$ radius leading to complete suppression of $j/\psi$ in quark-gluon plasma. For the current
experimental results on $j/\psi$ suppression/production at SPS, RHIC and LHC see \cite{sps,rhic,lhch}.
Note that the calculation done in \cite{satz} uses lattice QCD results at finite temperature in equilibrium
which may not be applicable at RHIC and LHC where the system may be in non-equilibrium.
This is because of the following reason.

Note that if the quark-gluon plasma thermalizes then the non-equilibrium stage can be called pre-equilibrium stage and the thermalized stage can be called equilibrium stage. However, since the two nuclei at RHIC and LHC travel almost at speed of light the longitudinal momenta of the partons inside the
nuclei just before the nuclear collision are much larger than their transverse momenta.
This leads to momentum anisotropy creating a non-equilibrium quark-gluon plasma just after the nuclear collisions.
Hence we know for sure that the quark-gluon plasma is formed in non-equilibrium at RHIC and LHC. Note that although the "non equilibrium" phase and  "equilibrium" phase may not exclude each other but, as mentioned above, we know for sure that the quark-gluon plasma is formed in non-equilibrium but we do not know for sure if the quark-gluon plasma is in equilibrium at RHIC and LHC. This is because sufficient number of secondary partonic collisions is necessary to bring the quark-gluon plasma to equilibrium which can happen if the life time of the QCD medium is large. However, there is no way of directly measuring the life time of the QCD medium because all we experimentally measure are hadrons and color singlet observables. Hence we do not know whether the life time of QCD medium is small or large. All we know is that the hadronization time scale in QCD is very small ($\sim 10^{-24}$ seconds). As mentioned in \cite{nayakjt} even if the experimental data at RHIC and LHC are explained by assuming thermalization, it does not prove that the quark-gluon plasma is thermalized. This is because in order to prove that the quark-gluon plasma is thermalized one has to prove that the same experimental data can not be explained by non-equilibrium quark-gluon plasma. This is because, as mentioned above, we know for sure that the quark-gluon plasma is formed in non-equilibrium at RHIC and LHC. In order to prove that the same experimental data can not be explained by non-equilibrium quark-gluon plasma one has to study non-equilibrium and non-perturbative QCD.

Note that when we say non-equilibrium (and factorization) that does not mean that there may not be any $j/\psi$ suppression due to Debye screening in non-equilibrium QGP. What we said is that the lattice QCD calculation is valid in equilibrium with static temperature $T$ and the lattice QCD calculation is not valid in non-equilibrium QCD where one can not define a temperature. As mentioned above when the Debye screening radius becomes smaller than the $j/\psi$ radius then there is complete suppression of $j/\psi$ in QGP \cite{satz}. The Debye screening radius and the $j/\psi$ radius were calculated in \cite{satz} by assuming equilibrium QGP at static temperature. Similarly, to observe the $j/\psi$ suppression in non-equilibrium QGP one has to calculate the Debye screening radius and the $j/\psi$ radius in non-equilibrium QCD which is not easy. For example when charm and anticharm are in color octet state, like that in NRQCD, the color potential is not Coulomb potential.
It should be mentioned here that the exact potential produced by the color charge of the quark is not known till now, see for example \cite{nayakj,nayake} for recent development of the general form of the color potential produced by the color charge of the quark. Hence the calculation of $j/\psi$ radius in non-equilibrium QCD in NRQCD color octet mechanism is not available as lattice QCD calculation is not applicable in non-equilibrium QCD. In addition to this, the non-perturbative Debye screening mass in non-equilibrium QCD is not available because lattice QCD is not applicable in non-equilibrium QCD although the Debye screening mass and the magnetic screening mass at one loop level in non-equilibrium QCD are available in \cite{cooper,cooper1,birse}. Hence unless we calculate the non-perturbative Debye screening mass and the $j/\psi$ radius in non-equilibrium QCD it is not possible to say if there is $j/\psi$ suppression via Debye screening in non-equilibrium QCD at RHIC and LHC or not.

The production of heavy quark-antiquark pair occurs in the hard scattering in the initial nuclear
collisions at RHIC and LHC. There can also be secondary heavy quark-antiquark pair production from parton
fusion processes from the QCD medium. The production of heavy quark-antiquark pair can be
calculated by using pQCD because the relevant momentum transfer scale $2M$ is large
where $M$ is the mass of the heavy quark.

The formation of heavy quarkonium from heavy quark-antiquark pair involves non-perturbative QCD
which is not solved yet. Hence we depend on experimental data to extract the non-perturbative matrix
element of heavy quarkonium production. This is in contrast to bound state formation in QED such as
the hydrogen atom or the positronium where the potential energy is known to be of the Coulomb form.
Although phenomenological potentials, like Coulomb potential or Coulomb plus linear potential etc.
\cite{singf,pot,bodwinpot}, are used in QCD to describe heavy quarkonium formation the exact form of the
potential energy between quark and antiquark is not known. This is because the exact form of the classical
Yang-Mills potential $A^{\mu a}(x)$ is not known yet. Hence the theoretical understanding
of bound state hadron formation in QCD remains a challenge.

In order to study heavy quarkonium production mechanism from heavy quark-antiquark pair in
non-equilibrium QCD it is necessary to understand the corresponding mechanism in
pp collisions at the same center of mass energy. NRQCD color octet mechanism for heavy quarkonium production
\cite{nrqcd} is widely used to explain experimental data at high energy colliders such as at Tevatron
\cite{tevatron} and LHC \cite{lhc}. In NRQCD the heavy quark-antiquark pair production in
color octet state is included and the non-perturbative NRQCD matrix elements are extracted from the experiments
\cite{oct}.

The PHENIX collaboration experimental
data of heavy quarkonium production in pp collisions at $\sqrt{s}$ = 200 GeV at RHIC
can be explained by using NRQCD color octet mechanism
\cite{nayakcooper,kniehl}. Similarly the ALICE
collaboration experimental data of heavy quarkonium production in pp collisions at $\sqrt{s}$ = 2.76 TeV at LHC
can be explained by using NRQCD color octet mechanism \cite{nrqcdalice}. Hence in order to compare the heavy quarkonium
production data in Au-Au collisions with pp collisions at $\sqrt{s}$ = 200 GeV at RHIC and in Pb-Pb collisions with
pp collisions at $\sqrt{s}$ = 2.76 TeV at LHC it may be necessary to study heavy quarkonium production in NRQCD color
octet mechanism in high energy heavy-ion collisions. As explained above, since the two nuclei at RHIC and LHC
collide almost at speed of light the quark-gluon plasma may be in non-equilibrium.
Hence it is necessary to study heavy quarkonium production in NRQCD color octet mechanism
in non-equilibrium QCD.

In order to study heavy quarkonium production at high energy colliders one needs to prove factorization
theorem, otherwise one will predict infinite cross section of heavy quarkonium
\cite{collinssterman, sterman, nayaksterman,nayaknr, nayakjt,nayakjp, nayaksterman1, nayakall}.
Note that in the original formulation of NRQCD color octet mechanism of heavy quarkonium production \cite{nrqcd}
the proof of factorization theorem was missing.
We have proved NRQCD factorization of heavy quarkonium production in color octet mechanism at NNLO in
coupling constant by using diagrammatic
approach in \cite{nayaksterman}. In \cite{nayaknr} we have proved NRQCD factorization of heavy
quarkonium production in color octet mechanism at all order in
coupling constant by using path integral formulation of QCD.

Note that heavy quarkonium is a primary signature of quark-gluon plasma detection \cite{satz}. Hence it is necessary
to prove factorization of heavy quarkonium production in non-equilibrium
QCD. In this paper we will prove factorization of heavy quarkonium production in non-equilibrium QCD
in (NRQCD) color octet mechanism at all order in coupling constant.

Note that in the formulation of NRQCD an ultraviolet cutoff $\Lambda \sim M$ is introduced \cite{nrqcd}. Hence the
ultraviolet (UV) behavior of NRQCD and QCD differ but infrared (IR) behavior of QCD and NRQCD remains same \cite{stewart}.
Because of this reason our analysis of infrared divergences in this paper is valid for QCD as well for NRQCD.

The main result we find in non-equilibrium QCD is that while the gauge links are not required in the
color singlet S-wave non-perturbative matrix elements in non-equilibrium QCD,
the gauge links are required in the S-wave color octet non-perturbative matrix elements in non-equilibrium QCD.
In the S-wave color singlet heavy quarkonium production the infrared divergences are absent
because the infrared divergences due to the soft gluons exchange between the heavy quark and
the nearby light-like quark (or gluon) in non-equilibrium QCD cancel with the corresponding infrared divergences due
to the soft gluons exchange between the heavy antiquark and the same nearby light-like quark
(or gluon).

However, there are uncanceled infrared divergences if the heavy quark-antiquark pair is in the
color octet state in non-equilibrium QCD and these uncanceled infrared divergences cancel with the corresponding infrared
divergences in the gauge links in the S-wave color octet non-perturbative matrix element
in non-equilibrium QCD at all order in coupling constant. We find that the (NRQCD) S-wave color octet
non-equilibrium and non-perturbative matrix element is independent of the light-like vector
$l^\mu$ which defines the gauge link. This proves factorization of heavy quarkonium production
in (NRQCD) color octet mechanism in non-equilibrium QCD at all order in
coupling constant.

The paper is organized as follows. In section II we briefly discuss the heavy quark-antiquark
pair in non-equilibrium QCD by using closed-time path integral formalism.
In section III we describe infrared divergences in heavy quarkonium production.
In section IV we prove factorization of heavy quarkonium production in non-equilibrium QCD
in (NRQCD) color octet mechanism at all order in coupling constant.
In section V we show that the factorization theorem is a key ingredient in calculation of
heavy quarkonium production cross section in non-equilibrium QCD. Section VI contains conclusions.

\section{ Heavy Quark-Antiquark Pair in Non-Equilibrium QCD Using Closed-Time Path Integral Formalism }

The ground state at RHIC and LHC due to the presence of QCD medium
at the initial time $t=t_{in}$ is not a vacuum state $|0>$. We denote the initial state
in non-equilibrium QCD at the initial time $t=t_{in}$ by $|in>$. We use the notation $\Psi$ for the
heavy quark field and the notation $\psi_l$ for the light quark field where $l=1,2,3=u,d,s$ stands
for up, down and strange quark respectively. The mass of the light quark is denoted by $m_l$ and
the mass of the heavy quark is denoted by $M$.

In the path integral formulation of QCD the non-equilibrium and non-perturbative
heavy quark-antiquark correlation function of the type
$<in|{\bar \Psi}_r(x_1) O_1\Psi_r(x_1) {\bar \Psi}_s(x_2)O'_2 \Psi_s(x_2)|in>$ is given by
\cite{muta,tucci,greiner,cooper}
\bea
&& <in|{\bar \Psi}_r(x_1) O_1 \Psi_r(x_1) {\bar \Psi}_s(x_2) O'_2 \Psi_s(x_2)|in>\nonumber \\
&&=\int \Pi_{n=1}^3 [d{\bar \psi}_{n+}][d{\bar \psi}_{n-}][d \psi_{n+} ][d\psi_{n-}][d{\bar \Psi}_{+}] [d{\bar \Psi}_{-}] [d \Psi_{+} ] [d\Psi_{-}][dQ_+] [dQ_-]\nonumber \\
&&{\bar \Psi}_r(x_1)  O_1 \Psi_r(x_1) {\bar \Psi}_s(x_2) O'_2 \Psi_s(x_2)
~{\rm det}(\frac{\delta \partial_\mu Q_+^{\mu c}}{\delta \omega_+^d})~~{\rm det}(\frac{\delta \partial_\mu Q_-^{\mu c}}{\delta \omega_-^d}) \nonumber \\
&& {\rm exp}[i\int d^4x [-\frac{1}{4}F^2[Q_+]+\frac{1}{4}F^2[Q_-]-\frac{1}{2 \alpha}(\partial_\mu Q_+^{\mu c })^2+\frac{1}{2 \alpha}(\partial_\mu Q_-^{\mu c })^2 \nonumber \\
&&+\sum_{l=1}^3[{\bar \psi}_{l+}  (\dslash[Q_+] -m_l)  \psi_{l+} -{\bar \psi}_{l-}  (\dslash[Q_-] -m_l)  \psi_{l-}]+{\bar \Psi}_{+}  (\dslash[Q_+] -M)  \Psi_{+}-{\bar \Psi}_{-}  (\dslash[Q_-] -M )  \Psi_{-}]]\nonumber \\
&& <Q_+,\psi_{u+},{\bar \psi}_{u+},\psi_{d+},{\bar \psi}_{d+},\psi_{s+},{\bar \psi}_{s+},\Psi_+,{\bar \Psi}_+,0|~\rho~|0,\Psi_-,{\bar \Psi}_-,{\bar \psi}_{s-},\psi_{s-},{\bar \psi}_{d-},\psi_{d-},{\bar \psi}_{u-},\psi_{u-},Q_-> \nonumber \\
\label{nezfi}
\eea
where $+(-)$ index corresponds to upper (lower) time branch in the closed-time path
formalism, $\rho$ is the initial density of state in non-equilibrium, $Q^{\mu c}(x)$ is the gluon field
with $c=1,2,...,8$ and
\bea
&& F^2[Q]={F}^a_{\mu \nu }[Q]{F}^{\mu \nu a}[Q],~~~~~~~~~~~~~~F^{\mu \nu a}[Q]=\partial^\mu Q^{\nu a}(x)-\partial^\nu Q^{\mu a}(x)+gf^{abc}Q^{\mu b}(x)Q^{\nu c}(x), \nonumber \\
&& \dslash[Q] =i\gamma^\mu \partial_\mu +gT^a\gamma^\mu Q^a_\mu.
\label{fmni}
\eea
The state
$|\Psi_\pm,{\bar \Psi}_\pm,{\bar \psi}_{s \pm },\psi_{s \pm },{\bar \psi}_{d \pm },\psi_{d \pm },{\bar \psi}_{u \pm },\psi_{u \pm },Q_\pm, 0>$
corresponds to the field configurations at the initial time $t=t_{in}=0$ where we work in the frozen ghost formalism
\cite{greiner,cooper} for the medium part at the initial time $t=t_{in}=0$. Note that the repeated closed-time path
indices $r,s=+,-$ are not summed.

For color singlet heavy quark-antiquark pair the operators $O_1$ and $O'_2$ are proportional to the unit matrix $I$ in color space
and for color octet heavy quark-antiquark pair the operators $O_1$ and $O'_2$ are proportional to the color matrix
$T^a$ in color space where $T^a$ is the generator of the SU(3) group.

\section{ Infrared Divergences in Heavy Quarkonium Production }

A detailed discussion of infrared divergences in heavy quarkonium production at high energy colliders is given in
\cite{nrqcd,nayaksterman,nayaknr}. As mentioned earlier the
ultraviolet (UV) behavior of NRQCD and QCD differ but infrared (IR) behavior of QCD and NRQCD remains same \cite{stewart}.
Hence the infrared divergences analysis in QCD is same as that in NRQCD.

For simplicity, let us consider the infrared divergences in QED before considering the infrared divergences
in QCD. For a real photon of four momentum $k^\mu$ emitted from an
incoming electron of four momentum $p^\mu$ we get by using the Feynman rules in QED the following contribution to the amplitude
\cite{grammer}
\begin{eqnarray}
&& \frac{1}{\pslash -\kslash -m}\not{\epsilon}(k)u(p)=-\frac{p \cdot \epsilon_{\rm pure}(k)}{k \cdot p}u(p)+\frac{\kslash \not{\epsilon}_{\rm phys}(k)}{2k \cdot p}u(p)
\label{totali}
\end{eqnarray}
where $m$ is the mass of the electron and
\begin{equation}
\not{\epsilon}(k)=\not{\epsilon}_{\rm phys}(k)+\not{\epsilon}_{\rm pure}(k),~~~~~~~~~~~~~~~~\not{\epsilon}_{\rm phys}(k) = \not{\epsilon}(k) -\kslash \frac{p \cdot \epsilon(k)}{k \cdot p},~~~~~~~~~~~~\not{\epsilon}_{\rm pure}(k) = \kslash \frac{p \cdot \epsilon(k)}{k \cdot p}\nonumber \\
\label{1ai}
\end{equation}
where $\epsilon_{\rm phys}^\mu(k)$ is the physical gauge field corresponding to the transverse polarization and
$\epsilon_{\rm pure}^\mu(k)$ is the pure gauge field corresponding to longitudinal polarization.

Infrared divergence occurs in QED due to the soft photon exchange in the limit $k^\mu \rightarrow 0$.
From eqs. (\ref{totali}) and (\ref{1ai}) we find
\begin{eqnarray}
&&\frac{p \cdot \epsilon_{\rm pure}(k)}{k \cdot p} \rightarrow \infty ~~~~~~~~~~~{\rm as}~~~~~~~~~~k^\mu \rightarrow 0, \nonumber \\
&&\frac{p \cdot \epsilon_{\rm phys}(k)}{k \cdot p} =0, \nonumber \\
&&  \frac{\kslash \not{\epsilon}_{\rm pure}(k)}{2k \cdot p} =0, \nonumber \\
&&\frac{\kslash \not{\epsilon}_{\rm phys}(k)}{2k \cdot p}  \rightarrow {\rm finite} ~~~~~~~~~~~{\rm as}~~~~~~~~~~k^\mu \rightarrow 0.
\label{totaldi}
\end{eqnarray}

\subsection{ Interaction of Non-Eikonal Current With the Gauge Field Generated by the Light-Like Eikonal Current in Quantum Field Theory }

The third equation $\frac{\kslash ~ \not{\epsilon}_{\rm pure}(k)}{2k \cdot p} =0$
of eq. (\ref{totaldi}) plays an essential role to use background field method
in quantum field theory to study factorization of infrared divergences due to the
presence of light-like Wilson line without modifying the finite value of the cross
section. This is because the gauge field generated
by the single light-like eikonal current can be replaced by a pure gauge
in quantum field theory in the situation in which that gauge field interacts with the
non-eikonal part of the diagram. This can be shown as follows.

From $\frac{1}{\pslash -\kslash -m}\not{\epsilon}(k)u(p)$ of eq. (\ref{totali})
we find that the non-eikonal part of the diagram gives $\frac{\kslash \not{\epsilon}(k)}{2k \cdot p}u(p)$.
Hence from the non-eikonal part of the diagram we find the relevant contribution
\bea
&& e\int \frac{d^4k}{(2\pi)^4} \frac{k^\nu \gamma_\nu \gamma_\mu A^\mu(k)}{2p \cdot k+i\epsilon}
= \int d^4x J_\mu(x) A^\mu(x)
\label{ggd}
\eea
where the non-eikonal current density $J^\mu(x)$ of the charge $e$ of four-momentum $p^\mu$ is given by
\bea
J^\mu(x) =\frac{e}{2} \gamma_\nu \gamma^\mu \int_0^{\infty} d\lambda \frac{\partial}{\partial x_\nu } \delta^{(4)}(x-p\lambda).
\label{nekcd}
\eea
Similarly from eq. (\ref{totali}) we find that the eikonal current density is given by \cite{nayaknr}
\bea
J^\mu(x) = e \int d\lambda ~l^\mu~\delta^{(4)}(x-l\lambda)
\label{ekcd}
\eea
where $l^\mu$ is the four-velocity of the eikonal current.

From the path integral formulation in quantum field theory we find that the effective action is given by \cite{nayaknr}
\bea
S_{eff}[J] = -\frac{1}{2} \int d^4x J^\mu(x)  \frac{1}{\partial^2}J_\mu(x).
\label{wj}
\eea
Hence using eqs. (\ref{nekcd}) and (\ref{ekcd}) in eq. (\ref{wj}) we find that the interaction between the non-eikonal current
and the gauge field generated by the light-like eikonal current in quantum field theory gives the effective (interaction)
action
\bea
&&{S}^{int}_{eff}[J] = \frac{e^2}{4}\int d^4x \gamma_\nu {\not l} \int_0^{\infty} d\lambda \delta^{(4)}(x-p\lambda) \frac{\partial}{\partial x_\nu }\frac{1}{\partial^2} \int d\lambda' ~\delta^{(4)}({ x}-l\lambda')\nonumber \\
&&= l^2 \frac{e^2}{2}\int d^4x [\frac{l \cdot \partial [p \cdot (x-p\lambda_0)]}{[p \cdot (x-p\lambda_0)]^2}][\frac{1}{(l \cdot x)^3}]
\label{effc}
\eea
where $\lambda_0$ is the solution of the equation
\bea
(x-p\lambda_0)^\mu (x-p\lambda_0)_\mu=0.
\label{l0}
\eea
From eqs. (\ref{effc}) and (\ref{l0}) we find that the interaction between the non-light-like non-eikonal current
and the gauge field generated by the light-like eikonal current in quantum field theory gives the effective (interaction)
lagrangian density
\bea
{\cal L}_{eff}^{int}(x) = l^2 \frac{e^2}{2}\frac{(p \cdot l) (p \cdot x) -(l \cdot x) p^2 }{(l \cdot x)^3[(p \cdot x)^2 -p^2 x^2]^{\frac{3}{2}}}.
\label{wjfp}
\eea
From eq. (\ref{wjfp}) we find that the interaction between the light-like non-eikonal current
and the gauge field generated by the light-like eikonal current in quantum field theory gives the effective (interaction)
lagrangian density
\bea
{\cal L}_{eff}^{int}(x) = \frac{e^2}{2} \frac{l^2 (p \cdot l) }{(p \cdot x)^2(l \cdot x)^3}.
\label{wjfpl}
\eea
For light-like four-velocity we get
\bea
l^2=l^\mu l_\mu=0.
\label{l2g}
\eea
Hence from eqs. (\ref{wjfp}), (\ref{wjfpl}) and (\ref{l2g}) we find that the effective (interaction) lagrangian density
due to the interaction between the (light-like or non-light-like) non-eikonal current and the gauge field generated by
the light-like eikonal current in quantum field theory is given by
\bea
{\cal L}_{eff}^{int}(x) =0,~~~~~~~{\rm when} ~~~~~~~l \cdot x \neq 0,~~~~~p \cdot x \neq 0.
\label{wjf}
\eea
Hence from eq. (\ref{wjf}) we find that the interaction between the non-eikonal current and the gauge field generated by the
light-like eikonal current in quantum field theory gives zero effective (interaction) lagrangian density. This is also evident
from the third equation $\frac{\kslash ~\not{\epsilon}_{\rm pure}(k)}{2k \cdot p} =0$
of eq. (\ref{totaldi}). This is obvious because pure gauge field corresponds to longitudinal polarization
which can not contribute to physical cross section. Similarly we have
shown in \cite{nayaknr} that the light-like eikonal current generates zero effective lagrangian density. Hence from
eq. (\ref{wjf}) and \cite{nayaknr} we find that
the light-like eikonal current in quantum field theory generates pure gauge field which agrees with the corresponding
result in classical mechanics \cite{collinssterman,nayakj,nayake}.

\subsection{ Infrared Divergences Due to the Presence of Light-Like Wilson Line }

Hence from eqs. (\ref{totaldi}), (\ref{1ai}), (\ref{wjf}), (\ref{totali}) and \cite{nayaknr}
we find that the non-eikonal part $\frac{\kslash ~\not{\epsilon}(k)}{2k \cdot p} $ of the diagram is
necessary to calculate the finite cross section but is not necessary to calculate the
relevant infrared divergences which can be calculated by using the eikonal part $\frac{p \cdot \epsilon(k)}{k \cdot p}$
of the diagram. In addition to this we find from eq. (\ref{totaldi}) that
the physical gauge field which corresponds to transverse polarization does not contribute to
infrared divergences in quantum field theory and the pure gauge field corresponding to the longitudinal polarization
does not contribute to the finite cross section in quantum field theory.

Since the eikonal current of the light-like charge generates pure gauge field in quantum field theory \cite{nayaknr} (see also eq. (\ref{wjf}) above)
we find from eqs. (\ref{wjf}), (\ref{totaldi}) and \cite{nayaknr} that the infrared divergences in QED due to the
soft-photons exchange with the the light-like Wilson line in QED can be studied by using pure gauge without
modifying the finite value of the cross section. Since the pure gauge field corresponds to unphysical longitudinal polarization,
it can be gauged away in the sense of factorization. Hence we find that the analysis of infrared divergences in quantum
field theory due to the presence of light-like Wilson line can be simplified by using pure gauge.

Note that the eikonal current of the light-like charge generates pure gauge field in classical
mechanics \cite{collinssterman,nayakj,nayake} and in quantum field theory \cite{nayaknr} (also see eq. (\ref{wjf}) above)
at all time-space position $x^\mu$ except at the position transverse to the motion of the
charge (${\vec l} \cdot {\vec x}=0$) at the time of closest approach ($x_0=0$).
We are interested in the infrared behavior of the non-perturbative matrix element of the
type $<in|{\bar \Psi}_r(x_1)  O_1 \Psi_r(x_1) {\bar \Psi}_s(x_2) O'_2 \Psi_s(x_2)|in>_A$ in the
presence of nearby light-like quark (or gluon). Since $<in|{\bar \Psi}_r(x_1)  O_1 \Psi_r(x_1) {\bar \Psi}_s(x_2) O'_2 \Psi_s(x_2)|in>$
is a non-perturbative matrix element its property at all order in coupling constant can be studied by using path integral
formulation of QCD. In addition to this it is clear that since the operators $O_1,O_2$ contain color matrices $T^a$
the non-perturbative matrix element \\
$<in|{\bar \Psi}_r(x_1)  O_1 \Psi_r(x_1) {\bar \Psi}_s(x_2) O'_2 \Psi_s(x_2)|in>_A$
in the presence of light-like Wilson line is not gauge invariant. Since $<in|{\bar \Psi}_r(x_1)  O_1 \Psi_r(x_1) {\bar \Psi}_s(x_2) O'_2 \Psi_s(x_2)|in>_A$ is not gauge invariant it can not cancel the uncanceled infrared divergences due to the interaction between
color octet heavy quark-antiquark pair and the nearby light-like quark (or gluon). Hence we find that both the issues of gauge
invariance and factorization of soft and collinear divergences can be resolved by using symmetry consideration
at the lagrangian level (instead of diagrammatic technique) by using path integral formulation of quantum
field theory in the background field method in the presence of pure gauge background field \cite{tucci}.

In QCD the SU(3) pure gauge is given by
\bea
T^aA^{\mu a}(x) = \frac{1}{ig} [\partial^\mu \Phi(x)]\Phi^{-1}(x)
\label{gtqcdi}
\eea
which gives the non-abelian gauge link \cite{nayaknr,nayakall,nayaknr}
\bea
\Phi(x)={\cal P}e^{-ig \int_0^{\infty} dt l\cdot { A}^c(x+lt) T^c }=e^{igT^c\omega^c(x)}
\label{ngli}
\eea
for infrared divergences due to infinite number of soft gluons exchange with the light-like quark
where ${\cal P}$ is the path ordering and $l^\mu$ is the four-velocity of the light-like quark.

From this point of view it
is useful to recall that the path integral formulation of background field method was first used by Tucci \cite{tucci}
to prove factorization of soft and collinear divergences by using symmetry consideration at the lagrangian
level rather than the diagrammatic approach. Since the path integral formulation of the background field method
of QED is more general than the path integral formulation of QED, the path integral formulation of the background
field method of QED to prove factorization of soft and collinear divergences in QED is based on first principle
calculation. The proof of factorization of soft and collinear divergences in the path integral formulation of
background method by Tucci in \cite{tucci} is extended to non-equilibrium QED in \cite{nayakqed}.

Note that the calculation done in \cite{tucci} is not an exact extension from QED to QCD. Let us clarify what we
meant when we say that the calculation done in \cite{tucci} is not an exact extension from QED to QCD. The main result of \cite{tucci} is eq. (4.18) which is obtained by using the generating functional $Z^{a,G+\Delta G}[{\cal J}]$ from eq. (4.17) using the background Feynman gauge fixing $G+\Delta G$ instead of the background field gauge fixing $G$ as given by eq. (2.2). For one soft gluon the diagrams are given by Fig. 3 of \cite{tucci}. As can be seen from this figure only the first two diagrams of the right hand side correspond to one soft photon case of QED obtained from eq. (1.6) of \cite{tucci}. In this sense the calculation of \cite{tucci} is not an exact extension from QED to QCD. If the generating functional $Z^{a,G}[{\cal J}]$ from eq. (2.2) in the background field gauge fixing is used [like in our study or in \cite{abbott,ab1}] instead of the generating functional $Z^{a,G+\Delta G}[{\cal J}]$ in the background Feynman gauge fixing from eq. (4.17) then the remaining diagrams (like the last two diagrams in the right hand side of Fig 3) will be absent and the result will be an exact extension of eq. (1.6) of QED to QCD. This can be mathematically seen as follows.

The eq. (4.18) is not similar to eq. (1.6) of QED in \cite{tucci} because of the presence of $R(x_2)$ and $R(x_1)$ in eq. (4.18) where $R(x)$ is given by eq. (4.15) which contains ghost field ($c$), gluon field ($Q$) and background pure gauge field $(A)$. Because of this, the right hand side of eq. (4.18) actually contains (full) correlation function of the type \\
$\int d^4y <c(x_2) {\bar c}(y) Q_\mu^a(y) \psi(x_2) {\bar \psi}(x_1)>~\Phi_{adjoint}(y)\partial^\mu \omega^b(y)$ etc. instead of $ <\psi(x_2) {\bar \psi}(x_1)>$ which appears in eq. (1.6). In this sense eq. (4.18) is not an exact extension of eq. (1.6) of QED to QCD because of the presence of these extra non-perturbative correlation functions of the type $<c(x_2) {\bar c}(y) Q_\mu^a(y) \psi(x_2) {\bar \psi}(x_1)>$ etc. This was expected because these additional terms involving $R(x_2)$ and $R(x_1)$ appeared because \cite{tucci} used background Feynman gauge fixing $G+\Delta G$ in eq. (4.17) instead of the background field gauge gauge fixing $G$ from eq. (2.2) to define the correlation function. It should be mentioned here that $<\psi(x_2) {\bar \psi}(x_1)>_A$ is not a physical observable quantity in QCD.

Note that the main advantage of the background field method of QCD (see \cite{abbott,ab1})
is that it retains explicit gauge invariance even after the gauge fixing term is added which greatly simplifies the calculation. This means the choice of gauge fixing term is severely restricted in order to maintain explicit gauge invariance which can greatly simplify the calculation. The background Feynman gauge fixing $G+\Delta G$ which is used in eq. (4.17) of \cite{tucci} is not gauge invariant. Hence the main advantage of using the background field method of QCD is lost in \cite{tucci} when the background Feynman gauge fixing term is used in eq. (4.17).

Note that a physical observable quantity in QCD is independent of the gauge fixing. Hence one will obtain the same physical observable quantity whether he/she works in background Feynman gauge fixing $G+\Delta G$ or in background field gauge fixing $G$. Therefore it is useful to choose a gauge fixing which greatly simplifies the calculation. The background field gauge fixing $G$ enormously simplifies the study of renormalization of ultra violet (UV) divergences in QCD because (unlike background Feynman gauge fixing $G+\Delta G$) the background field gauge fixing is explicitly gauge invariant, see \cite{abbott,ab1}. As is shown in \cite{nayaksterman} the factorization of infrared (IR) divergences and the gauge invariance are related. Hence one expects that the background field gauge fixing (which maintains explicit gauge invariance) will also enormously simplify the study of factorization of infrared (IR) divergences in QCD. This is what we find in our study in the background field gauge fixing where we obtain  $<\psi(x_2) {\bar \psi}(x_1)>_A=\Phi(x_2)<\psi(x_2) {\bar \psi}(x_1)>\Phi^\dagger(x_1)$ \cite{nayaknr,nayakall} which is an exact extension of eq. (1.6) of QED in \cite{tucci} to QCD because it does not contain extra $R(x_2)$ and $R(x_1)$ terms which are present in eq. (4.18).

By using the path integral formulation of the background field method of QCD in the background field gauge fixing we have proved factorization in QCD
at all orders in coupling constant in \cite{nayaknr,nayakall,nayaka2} by using the gauge fixing identity in QCD \cite{nayakgfi}. We have extended this to quark fragmentation function in non-equilibrium QCD in \cite{nayaka3}, to gluon fragmentation function in non-equilibrium QCD in \cite{nayakjp} and to $\chi_{cJ}$ production in non-equilibrium QCD in color singlet mechanism in \cite{nayakjt}. In this paper we will extend this to NRQCD heavy quarkonium production in non-equilibrium QCD in color octet mechanism. In \cite{nayaka2,nayaka3,nayakgfi} the proof was presented by using infinitesimal ($\omega^a$ small) gauge transformation technique. In this paper we present the proof by doing the exact finite gauge transformation calculation (see section IV).

\section{ Proof of Factorization of Heavy Quarkonium production in non-equilibrium QCD in (NRQCD) Color Octet Mechanism }

Note that the definition of the heavy quarkonium production matrix element from heavy quark-antiquark pair in
non-equilibrium QCD is a
non-perturbative quantity in non-equilibrium QCD which can not be calculated by using perturbation
theory in non-equilibrium QCD.
It is well known that a non-perturbative function can not be studied by using perturbation theory no matter
how many orders of perturbation theory is used. The path integral formulation (as opposed to diagrammatic methods in perturbation theory)
is necessary to study the properties of the non-perturbative matrix element of heavy quarkonium production in
non-equilibrium QCD at all order in coupling constant.

The non-equilibrium and non-perturbative heavy quark-antiquark correlation function of the type
$<in|{\bar \Psi}_r(x_1)  O_1 \Psi_r(x_1) {\bar \Psi}_s(x_2) O'_2 \Psi_s(x_2)|in>$ in the background field
method of QCD is given by \cite{greiner,cooper,tucci,thooft,zuber,abbott}
\bea
&& <in|{\bar \Psi}_r(x_1)  O_1 \Psi_r(x_1) {\bar \Psi}_s(x_2) O'_2 \Psi_s(x_2)|in>_A\nonumber \\
&&=\int \Pi_{n=1}^3 [d{\bar \psi}_{n+}][d{\bar \psi}_{n-}][d \psi_{n+} ][d\psi_{n-}][d{\bar \Psi}_{+}] [d{\bar \Psi}_{-}] [d \Psi_{+} ] [d\Psi_{-}][dQ_+] [dQ_-]\nonumber \\
&&{\bar \Psi}_r(x_1)  O_1 \Psi_r(x_1) {\bar \Psi}_s(x_2) O'_2 \Psi_s(x_2)
~{\rm det}(\frac{\delta G^c(Q_+)}{\delta \omega_+^d}) ~{\rm det}(\frac{\delta G^c(Q_-)}{\delta \omega_-^d}) \nonumber \\
&& {\rm exp}[i\int d^4x [-\frac{1}{4}F^2[Q_++A_+]+\frac{1}{4}F^2[Q_-+A_-]-\frac{1}{2 \alpha}(G^c(Q_+))^2+\frac{1}{2 \alpha}(G^c(Q_-))^2 \nonumber \\
&&+\sum_{l=1}^3[{\bar \psi}_{l+}  (\dslash[Q_++A_+] -m_l)  \psi_{l+} -{\bar \psi}_{l-}  (\dslash[Q_-+A_-] -m_l)  \psi_{l-}]\nonumber \\
&&+{\bar \Psi}_{+}  (\dslash[Q_++A_+] -M)  \Psi_{+}-{\bar \Psi}_{-}  (\dslash[Q_-+A_-] -M )  \Psi_{-}]]\nonumber \\
&&<Q_++A_+,\psi_{u+},{\bar \psi}_{u+},\psi_{d+},{\bar \psi}_{d+},\psi_{s+},{\bar \psi}_{s+},\Psi_+,{\bar \Psi}_+,0|~\rho~|0,\Psi_-,{\bar \Psi}_-,{\bar \psi}_{s-},\psi_{s-},{\bar \psi}_{d-},\psi_{d-},\nonumber \\
&&{\bar \psi}_{u-},\psi_{u-},Q_-+A_->
\label{cfqcdi}
\eea
where
\bea
&&F_{\mu \nu }^a[A_\pm +Q_\pm]=\partial_\mu [A_{\nu \pm}^a+Q_{\nu \pm}^a]-\partial_\nu [A_{\mu \pm}^a+Q_{\mu \pm}^a]+gf^{abc} [A_{\mu \pm}^b+Q_{\mu \pm}^b][A_{\nu \pm}^c+Q_{\nu \pm}^c], \nonumber \\
&&G^c(Q_\pm) =\partial_\mu Q^{\mu c}_\pm + gf^{cba} A_{\mu \pm}^b Q^{\mu a}_\pm=D_\mu[A_\pm]Q^{\mu c}_\pm
\label{gai}
\eea
and the type I gauge transformation is given by \cite{thooft,abbott,zuber}
\bea
&&T^cA'^c_{\mu \pm}(x)=U_\pm(x)T^cA^c_{\mu \pm}(x)U^{-1}_\pm(x)+ \frac{1}{ig}[\partial_\mu U_\pm(x)] ~U^{-1}_\pm(x),\nonumber \\
&&T^cQ'^c_{\mu \pm}(x)=U_\pm(x)T^cQ^c_{\mu \pm}(x)U^{-1}_\pm(x),~~~~~~~~~~~~~~~U_\pm(x)=e^{igT^c\omega^c_\pm(x)}.
\label{omegai}
\eea

When the operators $O_1,~O'_2$ are independent of quantum fields then
by changing the integration variable $Q \rightarrow Q-A$ in the right hand side of eq. (\ref{cfqcdi}) we find
\bea
&& <in|{\bar \Psi}_r(x_1)  O_1 \Psi_r(x_1) {\bar \Psi}_s(x_2) O'_2 \Psi_s(x_2)|in>_A\nonumber \\
&&=\int \Pi_{n=1}^3 [d{\bar \psi}_{n+}][d{\bar \psi}_{n-}][d \psi_{n+} ][d\psi_{n-}][d{\bar \Psi}_{+}] [d{\bar \Psi}_{-}] [d \Psi_{+} ] [d\Psi_{-}][dQ_+] [dQ_-]\nonumber \\
&&{\bar \Psi}_r(x_1)  O_1 \Psi_r(x_1) {\bar \Psi}_s(x_2) O'_2 \Psi_s(x_2)
~{\rm det}(\frac{\delta G^c_f(Q_+)}{\delta \omega_+^d}) ~{\rm det}(\frac{\delta G^c_f(Q_-)}{\delta \omega_-^d}) \nonumber \\
&& {\rm exp}[i\int d^4x [-\frac{1}{4}F^2[Q_+]+\frac{1}{4}F^2[Q_-]-\frac{1}{2 \alpha}(G^c_f(Q_+))^2+\frac{1}{2 \alpha}(G^c_f(Q_-))^2 \nonumber \\
&&+\sum_{l=1}^3[{\bar \psi}_{l+}  (\dslash[Q_+] -m_l)  \psi_{l+} -{\bar \psi}_{l-}  (\dslash[Q_-] -m_l)  \psi_{l-}]+{\bar \Psi}_{+}  (\dslash[Q_+] -M)  \Psi_{+}-{\bar \Psi}_{-}  (\dslash[Q_-] -M )  \Psi_{-}]]\nonumber \\
&&<Q_+,\psi_{u+},{\bar \psi}_{u+},\psi_{d+},{\bar \psi}_{d+},\psi_{s+},{\bar \psi}_{s+},\Psi_+,{\bar \Psi}_+,0|~\rho~|0,\Psi_-,{\bar \Psi}_-,{\bar \psi}_{s-},\psi_{s-},{\bar \psi}_{d-},\psi_{d-},{\bar \psi}_{u-},\psi_{u-},Q_->\nonumber \\
\label{cfqcd1i}
\eea
where
\bea
G^c_f(Q_\pm) =\partial_\mu Q^{\mu c}_\pm + gf^{cba} A_{\mu \pm}^b Q^{\mu a}_\pm - \partial_\mu A^{\mu c}_\pm=D_\mu[A_\pm]Q^{\mu c}_\pm-\partial_\mu A^{\mu c}_\pm
\label{gfai}
\eea
and eq. (\ref{omegai}) gives
\bea
&&T^cQ'^c_{\mu \pm}(x)=U_\pm(x)T^cQ^c_{\mu \pm}(x)U^{-1}_\pm(x)+ \frac{1}{ig}[\partial_\mu U_\pm(x)] ~U^{-1}_\pm(x),~~~~~~~U_\pm(x)=e^{igT^c\omega^c_\pm(x)}.\nonumber \\
\label{typeIi}
\eea
Eq. (\ref{cfqcd1i}) can be written as
\bea
&& <in|{\bar \Psi}_r(x_1)   O_1 \Psi_r(x_1) {\bar \Psi}_s(x_2) O'_2  \Psi_s(x_2)|in>_A\nonumber \\
&&=\int \Pi_{n=1}^3 [d{\bar \psi}'_{n+}][d{\bar \psi}'_{n-}][d \psi'_{n+} ][d\psi'_{n-}][d{\bar \Psi}'_{+}] [d{\bar \Psi}'_{-}] [d \Psi'_{+} ] [d\Psi'_{-}][dQ'_+] [dQ'_-]\nonumber \\
&&{\bar \Psi}'_r(x_1)  O_1 \Psi'_r(x_1) {\bar \Psi}'_s(x_2) O'_2 \Psi'_s(x_2)
~{\rm det}(\frac{\delta G^c_f(Q'_+)}{\delta \omega_+^d}) ~{\rm det}(\frac{\delta G^c_f(Q'_-)}{\delta \omega_-^d}) \nonumber \\
&& {\rm exp}[i\int d^4x [-\frac{1}{4}F^2[Q'_+]+\frac{1}{4}F^2[Q'_-]-\frac{1}{2 \alpha}(G^c_f(Q'_+))^2+\frac{1}{2 \alpha}(G^c_f(Q'_-))^2 \nonumber \\
&&+\sum_{l=1}^3[{\bar \psi}'_{l+}  (\dslash[Q'_+] -m_l)  \psi'_{l+} -{\bar \psi}'_{l-}  (\dslash[Q'_-] -m_l)  \psi'_{l-}]+{\bar \Psi}'_{+}  (\dslash[Q'_+] -M)  \Psi'_{+}-{\bar \Psi}'_{-}  (\dslash[Q'_-] -M )  \Psi'_{-}]]\nonumber \\
&&<Q'_+,\psi'_{u+},{\bar \psi}'_{u+},\psi'_{d+},{\bar \psi}'_{d+},\psi'_{s+},{\bar \psi}'_{s+},\Psi'_+,{\bar \Psi}'_+,0|~\rho~|0,\Psi'_-,{\bar \Psi}'_-,{\bar \psi}'_{s-},\psi_{s-}',{\bar \psi}'_{d-},\psi'_{d-},{\bar \psi}'_{u-},\psi'_{u-},Q'_->\nonumber \\
\label{cfqcd1vbi}
\eea
because a change of integration variables from unprimed variables to primed variables
does not change the value of the integration. Under the gauge transformation the quark fields
transform as
\bea
\psi'_{l \pm}(x)=e^{igT^c\omega^c_\pm(x)}\psi_{l \pm}(x),~~~~~~~~~~~~~~~~\Psi'_{ \pm}(x)=e^{igT^c\omega^c_\pm(x)}\Psi_{ \pm}(x).
\label{psii}
\eea
Since we are working in the frozen ghost
formalism for the medium part at the initial time \cite{greiner,cooper} the
$<Q_+,\psi_{u+},{\bar \psi}_{u+},\psi_{d+},{\bar \psi}_{d+},\psi_{s+},{\bar \psi}_{s+},\Psi_+,{\bar \Psi}_+,0|~\rho~|0,\Psi_-,{\bar \Psi}_-,{\bar \psi}_{s-},\psi_{s-},{\bar \psi}_{d-},\psi_{d-},{\bar \psi}_{u-},\psi_{u-},Q_->$ in eq. (\ref{nezfi}) corresponding to initial density of
state in non-equilibrium QCD is gauge invariant by definition. Hence from eqs. (\ref{typeIi}) and (\ref{psii}) we find \cite{nayaka3}
\bea
&&<Q'_+,\psi'_{u+},{\bar \psi}'_{u+},\psi'_{d+},{\bar \psi}'_{d+},\psi'_{s+},{\bar \psi}'_{s+},\Psi'_+,{\bar \Psi}'_+,0|~\rho~|0,\Psi'_-,{\bar \Psi}'_-,{\bar \psi}'_{s-},\psi_{s-}',{\bar \psi}'_{d-},\psi'_{d-},{\bar \psi}'_{u-},\psi'_{u-},Q'_->\nonumber \\
&&=<Q_+,\psi_{u+},{\bar \psi}_{u+},\psi_{d+},{\bar \psi}_{d+},\psi_{s+},{\bar \psi}_{s+},\Psi_+,{\bar \Psi}_+,0|~\rho~|0,\Psi_-,{\bar \Psi}_-,{\bar \psi}_{s-},\psi_{s-},{\bar \psi}_{d-},\psi_{d-},{\bar \psi}_{u-},\psi_{u-},Q_->.\nonumber \\
\label{nmmi}
\eea
When background field $A^{\mu a}(x)$ is the SU(3) pure gauge as given by eq. (\ref{gtqcdi}) then we find from
eqs. (\ref{typeIi}), (\ref{psii}), (\ref{gfai}) and (\ref{gtqcdi}) that \cite{nayaknr,nayakall,nayaknr}
\bea
&& [dQ'_\pm] =[dQ_\pm],~~~~~~~~~~~[d{\bar \psi}'_{l\pm}] [d \psi'_{l\pm} ]=[d{\bar \psi}_{l\pm}] [d \psi_{l\pm} ],~~~~~~~~~~~~[d{\bar \Psi}'_\pm] [d \Psi'_\pm ]=[d{\bar \Psi}_\pm] [d \Psi_\pm ],\nonumber \\
&&{\bar \psi}'_{l\pm} [i\gamma^\mu \partial_\mu -m_l +gT^c\gamma^\mu Q'^c_{\mu \pm}] \psi'_{l\pm}={\bar \psi}_{l\pm} [i\gamma^\mu \partial_\mu -m_l +gT^c\gamma^\mu Q^c_{\mu \pm}] \psi_{l\pm}, \nonumber \\
&&{\bar \Psi}'_\pm [i\gamma^\mu \partial_\mu -M +gT^c\gamma^\mu Q'^c_{\mu \pm}] \Psi'_\pm={\bar \Psi}_\pm [i\gamma^\mu \partial_\mu -M +gT^c\gamma^\mu Q^c_{\mu \pm}]\Psi_\pm,~~~~~~~~~{F}^2[Q'_\pm]={F}^2[Q_\pm] \nonumber \\
&& (G_f^c(Q'_\pm))^2 = (\partial_\mu Q^{\mu c}_\pm(x))^2,~~~~~~~~~~~{\rm det} [\frac{\delta G_f^c(Q'_\pm)}{\delta \omega^d_\pm}] ={\rm det}[\frac{ \delta (\partial_\mu Q^{\mu c}_\pm(x))}{\delta \omega^d_\pm}].
\label{gqp4ai}
\eea
Using eqs. (\ref{gqp4ai}), (\ref{nmmi}) and (\ref{psii}) in eq. (\ref{cfqcd1vbi}) we find
\bea
&& <in|{\bar \Psi}_r(x_1)   O_1 \Psi_r(x_1) {\bar \Psi}_s(x_2) O'_2  \Psi_s(x_2)|in>_A\nonumber \\
&&=\int \Pi_{n=1}^3 [d{\bar \psi}_{n+}][d{\bar \psi}_{n-}][d \psi_{n+} ][d\psi_{n-}][d{\bar \Psi}_{+}] [d{\bar \Psi}_{-}] [d \Psi_{+} ] [d\Psi_{-}][dQ_+] [dQ_-]\nonumber \\
&&{\bar \Psi}_r(x_1)  e^{-igT^c\omega^c_r(x_1)} O_1  e^{igT^c\omega^c_r(x_1)} \Psi_r(x_1) {\bar \Psi}_s(x_2)  e^{-igT^c\omega^c_s(x_2)}O'_2   e^{igT^c\omega^c_s(x_2)} \Psi_s(x_2) \nonumber \\
&& {\rm det}(\frac{\delta \partial_\mu Q_+^{\mu c}}{\delta \omega_+^d})~ {\rm det}(\frac{\delta \partial_\mu Q_-^{\mu c}}{\delta \omega_-^d}) ~ {\rm exp}[i\int d^4x [-\frac{1}{4}F^2[Q_+]+\frac{1}{4}F^2[Q_-]-\frac{1}{2 \alpha}(\partial_\mu Q_+^{\mu c })^2+\frac{1}{2 \alpha}(\partial_\mu Q_-^{\mu c })^2 \nonumber \\
&&+\sum_{l=1}^3[{\bar \psi}_{l+}  (\dslash[Q_+] -m_l)  \psi_{l+} -{\bar \psi}_{l-}  (\dslash[Q_-] -m_l)  \psi_{l-}]+{\bar \Psi}_{+}  (\dslash[Q_+] -M)  \Psi_{+}-{\bar \Psi}_{-}  (\dslash[Q_-] -M )  \Psi_{-}]]\nonumber \\
&& <Q_+,\psi_{u+},{\bar \psi}_{u+},\psi_{d+},{\bar \psi}_{d+},\psi_{s+},{\bar \psi}_{s+},\Psi_+,{\bar \Psi}_+,0|~\rho~|0,\Psi_-,{\bar \Psi}_-,{\bar \psi}_{s-},\psi_{s-},{\bar \psi}_{d-},\psi_{d-},{\bar \psi}_{u-},\psi_{u-},Q_->.\nonumber \\
\label{cfq5p2xi}
\eea
Using similar techniques as above we find
\bea
&& <in|{\bar \Psi}_r(x_1)  e^{igT^c\omega^c_r(x_1)} O_1  e^{-igT^c\omega^c_r(x_1)} \Psi_r(x_1) {\bar \Psi}_s(x_2)  e^{igT^c\omega^c_s(x_2)}O'_2   e^{-igT^c\omega^c_s(x_2)} \Psi_s(x_2)|in>_A\nonumber \\
&&=\int \Pi_{n=1}^3 [d{\bar \psi}_{n+}][d{\bar \psi}_{n-}][d \psi_{n+} ][d\psi_{n-}][d{\bar \Psi}_{+}] [d{\bar \Psi}_{-}] [d \Psi_{+} ] [d\Psi_{-}][dQ_+] [dQ_-]\nonumber \\
&&{\bar \Psi}_r(x_1)  O_1 \Psi_r(x_1) {\bar \Psi}_s(x_2) O'_2 \Psi_s(x_2)
~{\rm det}(\frac{\delta \partial_\mu Q_+^{\mu c}}{\delta \omega_+^d})~~{\rm det}(\frac{\delta \partial_\mu Q_-^{\mu c}}{\delta \omega_-^d}) \nonumber \\
&& {\rm exp}[i\int d^4x [-\frac{1}{4}F^2[Q_+]+\frac{1}{4}F^2[Q_-]-\frac{1}{2 \alpha}(\partial_\mu Q_+^{\mu c })^2+\frac{1}{2 \alpha}(\partial_\mu Q_-^{\mu c })^2 \nonumber \\
&&+\sum_{l=1}^3[{\bar \psi}_{l+}  (\dslash[Q_+] -m_l)  \psi_{l+} -{\bar \psi}_{l-}  (\dslash[Q_-] -m_l)  \psi_{l-}]+{\bar \Psi}_{+}  (\dslash[Q_+] -M)  \Psi_{+}-{\bar \Psi}_{-}  (\dslash[Q_-] -M )  \Psi_{-}]]\nonumber \\
&& <Q_+,\psi_{u+},{\bar \psi}_{u+},\psi_{d+},{\bar \psi}_{d+},\psi_{s+},{\bar \psi}_{s+},\Psi_+,{\bar \Psi}_+,0|~\rho~|0,\Psi_-,{\bar \Psi}_-,{\bar \psi}_{s-},\psi_{s-},{\bar \psi}_{d-},\psi_{d-},{\bar \psi}_{u-},\psi_{u-},Q_->. \nonumber \\
\label{cfq5p2i}
\eea
From eqs. (\ref{nezfi}) and (\ref{cfq5p2i}) we find
\bea
&&<in|{\bar \Psi}_r(x_1) O_1\Psi_r(x_1) {\bar \Psi}_s(x_2) O'_2 \Psi_s(x_2)|in>\nonumber \\
&&=<in|{\bar \Psi}_r(x_1) \Phi_r(x_1) O_1 \Phi^\dagger_r(x_1) \Psi_r(x_1) {\bar \Psi}_s(x_2) \Phi_s(x_2) O'_2 \Phi^\dagger_s(x_2) \Psi_s(x_2)|in>_A
\label{finalzibn}
\eea
where (see eq. (\ref{ngli}) and \cite{nayakall})
\bea
\Phi_r(x)={\cal P}e^{-ig \int_0^{\infty} dt l\cdot { A}^c_r(x+lt)T^c }.
\label{tttni}
\eea
Note that the creation operator of the hadron $a^\dagger_H$ is defined as follows
\bea
|H+X>=a^\dagger_H |X>,~~~~~~~~~~\sum_X|X><X|=1
\label{hd}
\eea
where $X$ stands for other final state hadrons. Since $a^\dagger_H$ is the creation operator of the hadron as
defined in eq. (\ref{hd}), one finds from eq. (\ref{hd}) that $a^\dagger_H$
is not a function of the quark fields $\psi_l(x),\Psi(x)$ and is not a function
of gluon field $Q^{\mu a}(x)$ which implies from eq. (\ref{finalzibn}) that
\bea
&&<in|{\bar \Psi}_r(x_1) O_1\Psi_r(x_1)a^\dagger_H a_H {\bar \Psi}_s(x_2) O'_2 \Psi_s(x_2)|in>\nonumber \\
&&=<in|{\bar \Psi}_r(x_1) \Phi_r(x_1) O_1 \Phi^\dagger_r(x_1) \Psi_r(x_1) a^\dagger_H a_H {\bar \Psi}_s(x_2) \Phi_s(x_2) O'_2 \Phi^\dagger_s(x_2) \Psi_s(x_2)|in>_A
\label{finalziv}
\eea
which proves factorization of infrared divergences at all order in coupling constant in non-equilibrium QCD.
Eq. (\ref{finalziv}) is valid in covariant gauge, in light-cone gauge, in general axial
gauges, in general non-covariant gauges and in general Coulomb gauge etc. respectively \cite{nayakall}.

The eq. (\ref{finalziv}) is exact extension of eq. (1.6) of \cite{tucci} of factorization in QED.

Note that the non-equilibrium and non-perturbative matrix elements \\
$<in|{\bar \Psi}_r(x_1) O_1\Psi_r(x_1) a^\dagger_H a_H{\bar \Psi}_s(x_2) O'_2 \Psi_s(x_2)|in>$ and \\
$<in|{\bar \Psi}_r(x_1) \Phi_r(x_1) O_1 \Phi^\dagger_r(x_1) \Psi_r(x_1)a^\dagger_H a_H {\bar \Psi}_s(x_2) \Phi_s(x_2) O'_2 \Phi^\dagger_s(x_2) \Psi_s(x_2)|in>_A$ in eq. (\ref{finalziv}) are obtained from the exact generating functionals.
Hence the eq. (\ref{finalziv}) is valid at all order in coupling constant in non-equilibrium QCD.

Hence we find that
$<in|{\bar \Psi}_r(x_1) \Phi_r(x_1) O_1 \Phi^\dagger_r(x_1) \Psi_r(x_1) a^\dagger_H a_H{\bar \Psi}_s(x_2) \Phi_s(x_2) O'_2 \Phi^\dagger_s(x_2) \Psi_s(x_2)|in>_A$
in eq. (\ref{finalziv}) is gauge invariant and is consistent with factorization of infrared divergences in non-equilibrium QCD
at all order in coupling constant.

From eq. (\ref{finalziv}) we find that the uncanceled infrared divergences due to the
interaction of the color octet heavy quark-antiquark pair with the nearby light-like quark (or gluon) in non-equilibrium QCD
cancel with the corresponding infrared divergences in the gauge links in the S-wave color octet non-perturbative matrix element
in non-equilibrium QCD at all order in coupling constant.
This proves factorization of infrared divergences of heavy quarkonium production
from color octet heavy quark-antiquark pair in non-equilibrium QCD at all order in
coupling constant.

As explained in \cite{nayakjp} the non-perturbative matrix element in non-equilibrium QCD can be
obtained by using $|Q>=a^\dagger |in>$ instead of $|Q>=a^\dagger |0>$ in vacuum where $a^\dagger$
is the creation operator of the heavy quark.
When the operators $O_1,~O'_2$ are proportional to the color matrix $T^a$ then we find from
eq. (\ref{finalziv}) that the $S-$wave color octet non-perturbative matrix element
for heavy quarkonium production in (NRQCD) color octet mechanism in non-equilibrium QCD is given by
\bea
<in|{\cal O}_H|in> = <in|\chi^\dagger(0) K_{n,e} \xi(0) \Phi_l^{(A)\dagger}(0)_{eb}(a^\dagger_H a_H) \Phi_l^{(A)}(0)_{ba}\xi^\dagger(0) K'_{n,a} \chi(0)|in>
\label{mnrqcdfactfi}
\eea
which is consistent with factorization of infrared divergences in non-equilibrium QCD
at all order in coupling constant where the gauge link
$\Phi_l^{(A)}$ in the adjoint representation of SU(3) is given by
\bea
\Phi_l^{(A)}(x)={\cal P}e^{-ig \int_0^{\infty} dt l\cdot { A}^c(x+lt)T^{(A)c} }
\label{adji}
\eea
where $T^{(A)c}_{ab}=-if^{cab}$. In eq. (\ref{mnrqcdfactfi}) $\xi$ is the two component Dirac spinor field that annihilates a heavy quark,
$\chi$ is the two component Dirac spinor field that creates a heavy quark, $a^\dagger_H$ is the creation
operator of the hadron and the operators $K_{n,e},K'_{n,a}$ are proportional to $T^e,T^a$ respectively.

Eq. (\ref{mnrqcdfactfi}) is similar to \cite{nayaksterman,nayaknr} except that the vacuum expectation is replaced by medium average.

Note that the non-equilibrium and non-perturbative matrix element \\
$<in|{\bar \Psi}_r(x_1) O_1\Psi_r(x_1) a^\dagger_H a_H{\bar \Psi}_s(x_2) O'_2 \Psi_s(x_2)|in>$
in the left hand side of eq. (\ref{finalziv}) is independent of $l^\mu$.
This proves that the long-distance behavior of the non-equilibrium and non-perturbative NRQCD matrix element
$ <in|\chi^\dagger(0) K_{n,e} \xi(0) \Phi_l^{(A)\dagger}(0)_{eb}(a^\dagger_H a_H) \Phi_l^{(A)}(0)_{ba}\xi^\dagger(0) K'_{n,a} \chi(0)|in>$
in eq. (\ref{mnrqcdfactfi}) is independent of the light-like vector
$l^\mu$ which defines the gauge link at all order in coupling constant.

This concludes the proof of factorization of heavy quarkonium production
in (NRQCD) color octet mechanism in non-equilibrium QCD at all order in
coupling constant.

\section{Factorization Theorem is a key ingredient in calculation of Heavy Quarkonium production cross section in non-equilibrium QCD}

In this section we will show how the factorization theorem as given by eqs. (\ref{finalzibn})
and (\ref{finalziv}) in non-equilibrium QCD
is actually a key ingredient in calculation of heavy quarkonium ($H$) production
cross section in color octet mechanism at all order in coupling constant in non-equilibrium QCD
by using the formula
\bea
d\sigma_{A+B \rightarrow H +X(P_T)} = d{\hat \sigma}_{A+B \rightarrow Q{\bar Q} +X(P_T)} ~<in|{\cal O}_{H}|in>
\label{cssn}
\eea
where $d{\hat \sigma}_{A+B \rightarrow Q{\bar Q} +X(P_T)}$ is the $Q{\bar Q}$ production cross section in color octet
state at all order in coupling constant in non-equilibrium QCD and $<in|{\cal O}_H|in>$ is the non-perturbative matrix
element of heavy quarkonium production in color octet mechanism in non-equilibrium QCD as given by eq. (\ref{mnrqcdfactfi}).

Let us prove how the eqs. (\ref{finalzibn}) and (\ref{finalziv}) are key ingredients
to prove eq. (\ref{cssn}) to calculate the heavy quarkonium production cross section in non-equilibrium QCD
in color octet mechanism at all order in coupling constant.
Suppose we calculate the
cross section $d{\hat \sigma}_{A+B \rightarrow Q{\bar Q} +X(P_T)}$ of heavy quark-antiquark
production in color octet state in non-equilibrium QCD at all order in coupling constant
in the presence of light-like quark (or gluon). Then from eq. (\ref{finalzibn}) we find
\bea
&&<in|{\bar \Psi}_r(x_1) O_1\Psi_r(x_1) {\bar \Psi}_s(x_2) O'_2 \Psi_s(x_2)|in>_A\nonumber \\
&&=<in|{\bar \Psi}_r(x_1) \Phi^\dagger_r(x_1) O_1 \Phi_r(x_1) \Psi_r(x_1) {\bar \Psi}_s(x_2) \Phi^\dagger_s(x_2) O'_2 \Phi_s(x_2) \Psi_s(x_2)|in>.
\label{finalzibnst}
\eea
Hence from eq. (\ref{finalzibnst}) we find that the $A$ dependence which arises due to the soft gluon exchanges with
light-like quark (or gluon) is factorized and only appears in the gauge-links $\Phi(x)$ in the right hand side where
$\Phi(x)$ is given by eq. (\ref{tttni}). Eq. (\ref{finalzibnst}) implies that in the cross section
$d{\hat \sigma}_{A+B \rightarrow Q{\bar Q} +X(P_T)}$ for $Q{\bar Q}$ production in color octet state
in non-equilibrium QCD at all order in coupling constant the infrared divergences due to the presence of
light-like quark (or gluon) are factorized only to the gauge links $\Phi(x)$.

Eq. (\ref{finalzibnst}) is the exact extension of eq. (1.6) of \cite{tucci} of the factorization in QED.

Hence from eq. (\ref{finalzibnst}) we find that the non-perturbative matrix element of heavy quarkonium production
in non-equilibrium QCD in color octet mechanism which cancels these
infrared divergences and is consistent with the factorization theorem
is obtained from eq. (\ref{finalziv}) and is given by eq. (\ref{mnrqcdfactfi}). This proves
that the factorization theorem as given by eqs. (\ref{finalzibn})
and (\ref{finalziv}) in non-equilibrium QCD is actually a key ingredient to prove eq. (\ref{cssn}) to calculate the
heavy quarkonium production cross section in color octet mechanism at all order in coupling constant in non-equilibrium
QCD.

\section{Conclusions}
Recently we have proved the NRQCD factorization in heavy quarkonium production at high energy colliders at all orders in coupling
constant. In this paper we have extended this to non-equilibrium QCD and have proved the NRQCD factorization in heavy quarkonium
production at all orders in coupling constant in non-equilibrium QCD. We have predicted the correct definition of the non-equilibrium and non-perturbative
NRQCD matrix element of heavy quarkonium production. We have shown that the correct definition of the non-equilibrium and non-perturbative
NRQCD matrix element of heavy quarkonium production is independent of the light-like four-velocity $l^\mu$ used to define the light-like gauge link.

As mentioned earlier the proof of factorization is to ensure that any remaining infrared (IR) divergence in the heavy quark and heavy antiquark production (along with light partons) is canceled with the correct definition of the non-perturbative NRQCD matrix element of the heavy quarkonium production at all orders in coupling constant in QCD. If such cancelation does not happen then one will predict infinite cross section of the heavy quarkonium production. Hence the proof of factorization in quantum field theory is an essential ingredient to measure heavy quarkonium production at high energy colliders and at high energy heavy-ion colliders. This is similar to renormalization in quantum field theory which deals with ultra violet (UV) divergences.

Note that the finite value of the cross section can be calculated by using the non-eikonal part as given by the last line of eq. (\ref{totaldi}) but its calculation is not necessary if we are studying the infrared (IR) divergence behavior due to the presence of light-like eikonal line which can be seen from the third line of eq. (\ref{totaldi}), see also \cite{nayaknr,nayaksterman}. Hence the infrared (IR) divergence behavior due to the presence of light-like eikonal line can be studied by using the eikonal approximation as given by the first line of eq. (\ref{totaldi}) without modifying the finite value of the cross section which can be seen from the third line of eq. (\ref{totaldi}), see also eq. (\ref{wjf}) and the discussion in the last paragraph of section 7 of \cite{nayaknr} for details. Hence the proof of factorization of infrared (IR) divergences in non-equilibrium QCD which we have presented in this paper only deals with the infrared (IR) divergences behavior without calculating the finite value of the heavy quarkonium production cross section. This implies that the proof of factorization in non-equilibrium QCD which we have presented in this paper will not tell us any thing about the effect of QGP on the finite value of the heavy quarkonium production cross section. The proof of factorization which we have presented in this paper is to make sure that one does not predict infinite cross section of heavy quarkonium production.

Note that when we say non-equilibrium QCD, it has both vacuum part and medium part. As far as Debye screening argument for $j/\psi$ suppression is concerned, as mentioned in the introduction, if the Debye screening radius is smaller than the $j/\psi$ radius then the $j/\psi$ will be completely suppressed in QGP \cite{satz}. However, we have not observed complete suppression of $j/\psi$ at RHIC and LHC. This implies that the study of the heavy quarkonium production mechanism in non-equilibrium QCD is necessary. Hence the proof of factorization of heavy quarkonium production in non-equilibrium QCD at all orders in coupling constant which we have presented in this paper is necessary to study the heavy quarkonium production at RHIC and LHC.

\end{document}